\documentclass[a4paper,aps,pra,noshowpacs,superscriptaddress,nofootinbib,preprint]{revtex4-1}
\usepackage[utf8]{inputenc}
\usepackage[T1]{fontenc}
\usepackage[UKenglish]{babel}
\usepackage{lmodern}
\usepackage[colorlinks]{hyperref}
\usepackage{graphicx}
\usepackage{amsmath,amssymb,amsthm}
\usepackage{xspace}
\usepackage{mathtools}
\usepackage{dsfont}
\usepackage{xcolor}
\usepackage{paralist}
\usepackage{array}
 
\usepackage{cancel}

\DeclareMathOperator{\tr}{Tr}

\newcommand{\id}{\mathds{1}}

\newcommand{\bra}[1]{\left\langle #1 \right|}
\newcommand{\ket}[1]{\left| #1 \right\rangle}

\newcommand{\ketbra}[2]{\left|#1\middle\rangle\middle\langle#2\right|}
\newtheorem*{theorem*}{Theorem}

\usepackage[normalem]{ulem} %need for strikethrough
\usepackage{cancel} % strike-out in mathmodecyan

\begin{document}
 
\title{Quantum Markovianity as a supervised learning task}
\author{Sally Shrapnel}
\email{s.shrapnel@uq.edu.au}
\affiliation{Centre for Engineered Quantum Systems, School of Mathematics and Physics, The University of Queensland, St Lucia, QLD 4072, Australia}
\author{Fabio Costa}
\affiliation{Centre for Engineered Quantum Systems, School of Mathematics and Physics, The University of Queensland, St Lucia, QLD 4072, Australia}
\author{Gerard Milburn}
\affiliation{Centre for Engineered Quantum Systems, School of Mathematics and Physics, The University of Queensland, St Lucia, QLD 4072, Australia}

\date{\today}
\begin{abstract}
Supervised learning algorithms take as input a set of labelled examples and return as output a predictive model. Such models are used to estimate labels for future, previously unseen examples drawn from the same generating distribution. In this paper we investigate the possibility of using supervised learning to estimate the dimension of a non-Markovian quantum environment. Our approach uses an ensemble learning method, the Random Forest Regressor, applied to classically simulated data sets. Our results indicate this is a promising line of research.
\end{abstract}

\maketitle

\section{Introduction}	

Predicting particular aspects of future phenomena is a task central to science. Medical practitioners gather physiological data to make predictions about the likelihood of future pathology. Meteorologists accumulate and analyse environmental data to make predictions about tomorrow's weather. In almost all fields of human endeavour, one faces the task of predicting the value of a future variable, based on observations about a range of past variables. 

Traditionally, scientists have approached such tasks from the perspective of models and theoretical constructs particular to their own field. More recently, an alternative approach has emerged. Developments in computer hardware, computer science and artificial intelligence have led to the possibility of machines that can learn to predict future phenomena  with impressive accuracy and efficiency. Early machine learning approaches involved hand-crafted solutions, tailored to perform optimally on specific tasks. Recent techniques more often adopt a `black box' approach, where `off-the-shelf' architectures leverage the high computing power of today's technology. Similar algorithms are now able to solve a large variety of scientifically dissimilar tasks, with the algorithm remaining ignorant of any domain-specific theoretical suppositions. 

The modern field of machine learning now provides a large suite of tools that can be applied to a wide variety of learning tasks.  Although there are a wealth of past  theoretical computer science results relating to the performance of these methods, advances are currently driven largely by direct experimental application.  As techniques are applied to new domains, a richer understanding of both the relevant search space and the limitations of particular approaches is gained. Utilising classical machine learning techniques to better understand the structure of quantum data is likely no exception. The application of machine learning to the quantum domain has only recently begun to receive attention~\cite{biamonte2017, dunjko2017}, and it is likely the benefits and pitfalls of specific approaches will become clearer with time. 

In this paper we apply an ensemble supervised classification technique, Random Forest Regression, to address a specific quantum information problem: classifying an unknown environment, interacting with a quantum system, as either Markovian or non-Markovian. We also consider, in the case of a non-Markovian environment, the task of estimating its dimension. Our definitions of Markovianity and non-Markovianity follow recent work on quantum Markovian processes~\cite{Costa2016,  giarmatzi2018quantum, Pollock2018, Pollock2018a, milz2016reconstructing}. These approaches use a process matrix approach to define conditions for Markovianity, and have been shown to unify previous approaches.

For many quantum-technology applications the presence of non-Markovian noise is difficult to characterise. The complexity grows with the dimension of the environment that needs to be modelled: reliable and efficient methods to detect and measure non-Markovianity would certainly be of use. Current approaches require full tomography of a multi-time process, necessitating multiple non-destructive measurements on the system \cite{Costa2016, Pollock2018a, giarmatzi2018quantum}. Here we consider the possibility of characterising non-Markovianity in a more practically accessible scenario to explore the possibility of learning information about the environment when one does not have access to tomographically complete information. Specifically, we consider a situation where the system of interest is subjected to a sequence of controlled unitary transformations, and a single measurement is performed at the end. 

Our results demonstrate that it is indeed possible to train a learning model to provide a good estimate of the dimension of a non-Markovian environment from the statistics of the final measurement.  Our approach represents first steps towards finding a practical solution to the problem of estimating non-Markovian noise, and suggests machine learning techniques may well prove useful in this context.

The structure of the paper is as follows: in Section 2 we introduce the problem and clarify the nature of the particular learning task we wish to solve. In Section 3, we introduce the reader to machine learning and detail the main technique used,  Random Forest Regression. In Section 4 we present the specific methods used and in Section 5 we present our results. We finish with a discussion.

\section{Quantum non-Markovianity}
We consider the following scenario: an experimentalist attempts to control and manipulate a system $S$, which can interact with some inaccessible environment $E$. For concreteness, we take $S$ to be a two-level system, while $E$ can in principle have arbitrary dimension. The experimentalist can perform operations on the system at some prescribed times $t_1,\dots,t_n$. The operation at time $t_n$ is restricted to be a projective measurement on one of the states
\begin{equation} \label{paulibases}
\ket{\psi_1}= \ket{x_+},\qquad \ket{\psi_2}= \ket{y_+},\qquad \ket{\psi_3}= \ket{z_+}.
\end{equation}
At the times $t_1,\dots,t_{n-1}$, the experimentalist can apply one of the unitary transformations
\begin{equation}
U_0 = \id,\qquad
U_1 = \sigma_x,\qquad
U_2 = \sigma_y,\qquad
U_3 = \sigma_z.
\label{pauliunis}
\end{equation}
(We assume that the time it takes to perform unitaries and measurements is short with respect to any other relevant dynamics, so each operation can effectively be considered as instantaneous.)
The experimenter thus collects statistics in the form of probabilities
\begin{equation}
p_{i_1\dots i_n} = P(\psi_{i_n}|U_{i_1},\dots, U_{i_{n-1}}).
\label{probabilities}
\end{equation}

Although $S$ can be coupled with an arbitrarily large environment, a subsystem of dimension $d^{2n-1}$ is sufficient to reproduce an arbitrary multi-time expression such as Eq.~\eqref{probabilities} \cite{Gutoski2006, Chiribella2009}; we shall identify the environment $E$ with such a subsystem without loss of generality. The observed probabilities~\eqref{probabilities} can be obtained by alternatively evolving an initial system-environment state, possibly correlated, with the controlled unitaries $U_{i_j}$ and some joint system-environment unitary evolution. This can be hard to model in general, both theoretically and computationally, given the exponential scaling of the relevant environment's dimension. The task of the experimenter is thus to estimate whether a simplification is possible, i.e., if it is possible to reproduce the observed statistics with a lower-dimensional effective environment.

\begin{figure}[ht]%
\includegraphics[width=0.75\columnwidth]{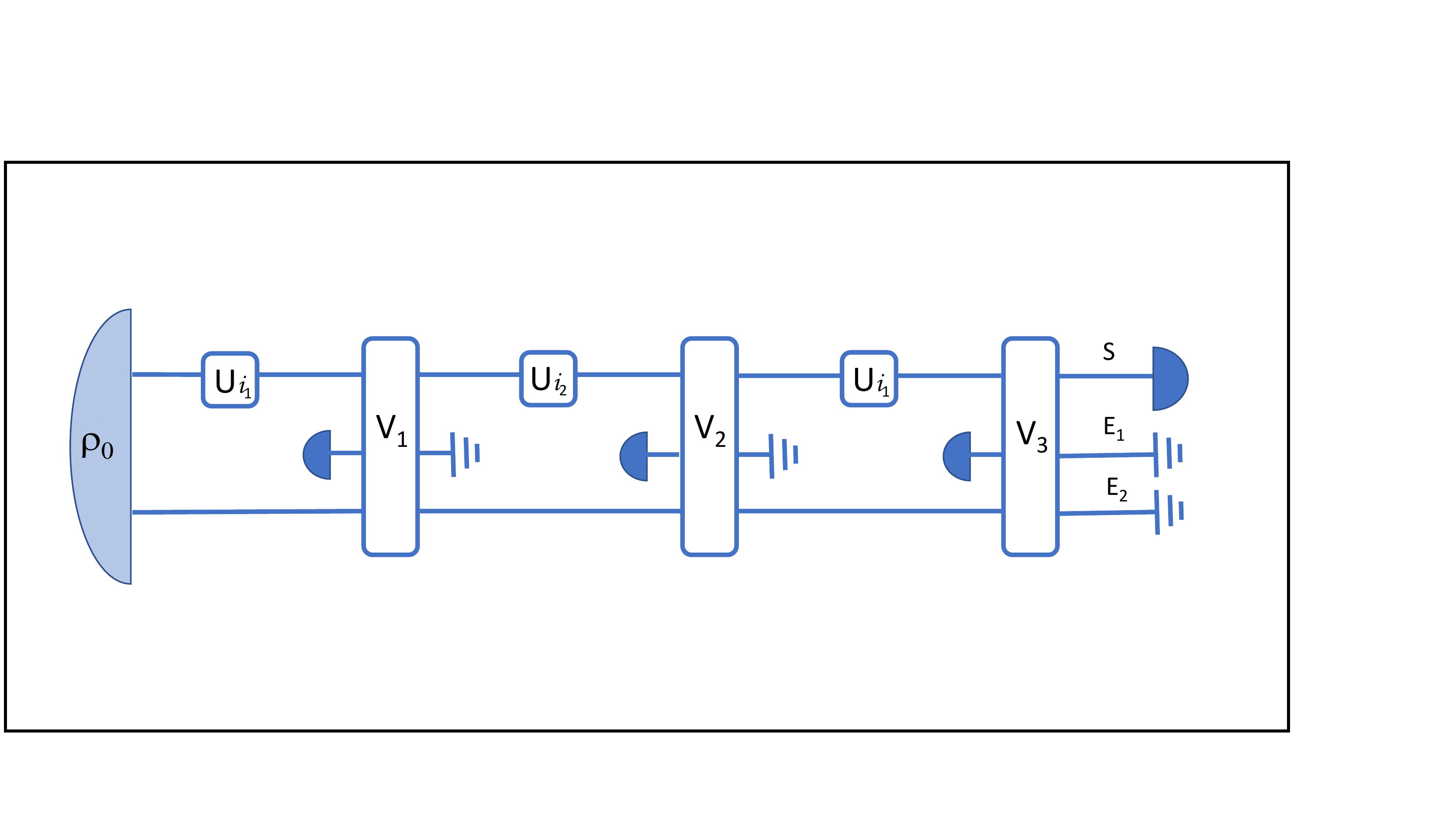}%
\caption{A circuit depiction of the experimental situation we consider. A system, $S$, is subjected to a series of controlled operations $U_{i_j}$ followed by a final projective measurement. The aim of our algorithm is to learn to estimate the dimension, $k_2$ of the non-Markovian part of the environment, $E_2$, that interacts with the system $S$ during the experimental procedure,  from observed measurement statistics alone. }%
\label{process}%
\end{figure}

More specifically, here we are interested in estimating how much of the environment carries memory of the system. We can formalise this question by decomposing $E$ into two subsystems at each time step: $E_1$ of dimension $k_1$ (the \emph{Markovian} environment) and $E_2$ of dimension $k_2$ (the \emph{non-Markovian} environment). After each time $t_j$, $1\leq j \leq n-1$, the Markovian environment is discarded and replaced with some fiducial state $\ket{0}$. Then the experimentally-controlled unitary $U_{i_j}$ is applied on the system $S$, after which system and environment (both Markovian and non-Markovian) evolve through a joint unitary $V_j$. Since the Markovian environment is discarded after each time step, we only need to explicitly keep track of the evolution of the system and the non-Markovian environment. The $SE_2$ state $\rho_j$ just before each time $t_{j+1}$ is then given by the recursive mapping
\begin{align}\label{step}
\rho_{j} &= \mathcal{E}_j \left(U_{i_j} \rho_{j-1}  \, U_{i_j}^{\dag}\right) \\
\label{dilated}
\mathcal{E}_j \left(\rho\right) :&=\tr_{E_1} \left[ V^{SE}_j \rho^{S E_2}\otimes \ketbra{0}{0}^{E_1} V^{S E\dag}_j \right].
\end{align}
where in Eq.~\eqref{step} the unitary $U_{i_j}$ is implicitly extended to act as identity on $E_2$, the superscripts on the right-hand-side of Eq.~\eqref{dilated} denote the subsystems on which the corresponding operators are defined, $\tr_X$ denotes the partial trace over a subsystem $X$, and $\rho_0$ is some initial joint state of system and non-Markovian environment, possibly correlated. The observed probabilities are finally given by
\begin{equation}
P(\psi_{i_n}|U_{i_1},\dots, U_{i_{n-1}}) = \bra{\psi_{i_n}} \tr_{E_2} \rho_{n-1} \ket{\psi_{i_n}},
\label{observed}
\end{equation}
where the dependence on the unitaries $U_{i_1},\dots, U_{i_n}$ on the right-hand-side is implicit in the state $\rho_{n-1}$ (Fig~\ref{process}).

If $k_2=1$, the whole process can be described by a sequence of trace-preserving maps acting on the system alone, concatenated with the experimentally-controlled operations. We say in this case the process is Markovian. A simple task is to estimate whether or not a system is Markovian, i.e., to verify whether $k_2$ is sufficiently close to $1$ within some confidence interval. This can be useful, for example, in a context where some protocol is only guaranteed to work in the presence of Markovian noise. More generally, the experimenter might be interested in estimating $k_2$ for a non-Markovian environment. This could be useful to evaluate the resources required to reproduce or simulate a process of interest, or to bound the space of processes to use for the analysis and optimisation of the experiment. Recall, in the most general case the dimension of the non-Markovian environment can be exponentially large in the number of time steps, $k_2=d^{2n-1}$ (see also~\cite{Gross}, particularly figure 5). Thus, finding a small $k_2$ can dramatically reduce the cost of modelling the process.

Therefore, the task we consider here for the experimenter is to estimate the smallest value of $k_2$ that correctly reproduces all possible observations on the system. Note that, for a given value of $k_2$, the amount of non-Markovian noise imparted on the system will vary depending on the strength of the interaction. Here we assume that such a strength, or the type of interaction, is not known by the experimenter, who attempts to deduce $k_2$ from the observed statistics only. We will assume that the strength and type of interaction with the environment is constant over time; we will come back to the details of the model in Section~\ref{datagen}.  

The process described above, where an environment retains memory of a system that can be probed at multiple times, is known as a \emph{channel with memory}~\cite{Kretschmann2005}, and it has also been studied in the context of quantum strategies \cite{Gutoski2006} and quantum networks \cite{Chiribella2008}. 
The corresponding notion of Markovianity we adopt agrees with that of quantum stochastic processes \cite{ACCARDI1978226, lindblad1979, Pollock2018} and quantum causal modelling \cite{Costa2016, Allen2016}. 
Note that, although different definitions of Markovianity exist in the literature \cite{li2017concepts}, approaches that characterise open dynamics in terms of a (time-dependent) map from an initial to a final state of the system \cite{Rivas2014} fall short of describing a scenario where multiple interventions at different times are possible within a single experimental run (as is the case for the situation we describe here).

%For a $d$-dimensional system evolving through $n$ time steps, an environment of total dimension $k_1 k_2\leq d^{2n-1}$ is sufficient to reproduce an arbitrary process [Watrous-Gutoski, Chiribella](with $k_1=1$ and $k_2=d^{2n-1}$ in the worst-case scenario). However, one would expect that a smaller non-Markovian dimension $k_2$ should be sufficient to capture several physically-meaningful scenarios. 

One possible method to estimate $k_2$ would be to fully reconstruct the multi-time process (i.e., the channel with memory) through tomography; this would however require performing measurements, and not just unitary transformations, at each time step \cite{Costa2016, Pollock2018a, giarmatzi2018quantum}. Therefore, one would expect that having access only to the probabilities \eqref{observed} would \emph{not} be sufficient to determine $k_2$ in general, because they range over settings that are not tomographically complete~\cite{milz2016reconstructing}. However, it remains unknown whether alternative methods exist that might permit one to estimate $k_2$ with a high probability of success for a broad range of physical situations.

Our approach is to cast this problem as a supervised learning task. We generate training data according to our knowledge of interacting quantum systems, where we label each data example by the values of the parameters used to generate that particular example. Specifically, our aim is to train the learning model to estimate the value of $k_2$ with low error on both training, validation and test set data. 

\section{Random Forests}

Understanding any machine learning application requires familiarity with some standard terminology. A computer program may be considered to `learn'  from an ``experience $E$ with respect to some class of tasks $T$ and performance measure $P$, if its performance in tasks in $T$, as measured by $P$, improves with experience $E$"~\cite{mitchell97}. Tasks are typically described according to how the learning system should process an \emph{example}, which is a collection of \emph{features}: quantitative values  measured from the system of interest. Each example is represented as a vector $x \in R^n$, where each vector entry $x_i$ corresponds to a specific feature. The learning experience $E$ is usually a dataset formed by a collection of independently and identically distributed examples (alternatively \emph{data points}). 

The algorithm is classed as \emph{supervised} when the experience $E$ includes a dataset where each example is also associated with a \emph{label} or \emph{target} (assumed to be provided with perfect accuracy by an expert ``supervisor"). The underlying assumption is that the output target variable does not take its value at random, but rather a relationship exists between the examples and their labels. Roughly speaking, the task of the model is to learn a function that maps examples to labels; a good approximation of the mapping function will permit accurate prediction of the value of an unseen output label, given the value of its input example. The performance measure $P$ is to some extent determined by the learning task: when one wishes to estimate the value of a categorical variable, e.g. ``has disease/does not have disease", the performance is judged according to the proportion of examples for which the model produces the correct output (the \emph{accuracy}). For regression tasks, where one wishes to estimate the value of a real variable, e.g. ``sale price", the performance is often judged by mean squared error (the error decreases as the Euclidean distance between model predictions and targets decreases).

The ultimate aim of machine learning is to perform optimally on new, previously unseen examples. Clearly, it is no good to perform perfectly on the data used to train the model (the `training data set' ), and perform poorly on unobserved examples. The ability for models to perform well on unseen data is called \emph{generalisation}. Common to all machine learning tasks is the problem of \emph{over-fitting}, where one fits the training data very well, but at the cost of generalisation. It is possible to estimate the generalisation error of a model by measuring the performance on a \emph{test set} of previously unseen examples. Thus the overall aim of machine learning is to not only minimise the training error, but  to also minimise the difference between the training and test error.

Random Forests (RFs) are a popular supervised learning method that require very few statistical assumptions about the underlying data, are easy to use, and have been applied with success in many domains~\cite{Breiman2001,biau2016random, louppe2014}. They provide an opportunity for parallel computation, have very few hyper-parameters to tune, and perform well on both high and low dimensional learning tasks\cite{Caruana08}. Consequently, Random Forests have enjoyed widespread use in a variety of scientific and industry related domains.  

\begin{figure}[t]%\vspace{-15pt}
\centerline{\includegraphics[width=1\columnwidth]{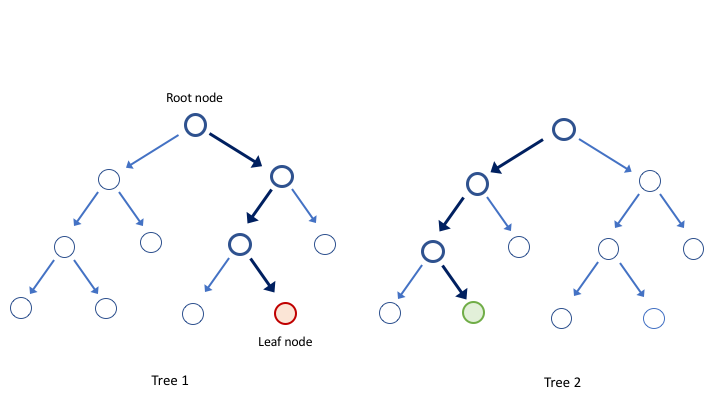}}
\vspace*{8pt}
\caption{Decision trees. Each decision tree starts by receiving an input to the root node corresponding to values for a random subset of features from a single example. Each subsequent branch is formed by procesing these values according to a thresholding decision. The final leaf node corresponds to a predicted label value (the labels are the value for $\log_2 k_2$ for the data we analyse here). Each of the examples is processed in this way.}
\label{twotrees}
\end{figure}

RFs are an example of ensemble learning - a class of algorithms that generate multiple classification or regression models and then aggregate their results.\footnote{There are a number of alternative ensemble methods that perform similarly to RFs across a variety of learning tasks. AdaBoost, Gradient Boosting Trees and XGBoost are all likely to provide models with similar success, though some studies show RFs have a slight advantage for data sets of high dimension~\cite{Caruana08}.} The basic unit is the \emph{decision tree}, a non-parametric model that is built by recursively partitioning the data space according to thresholding decisions (e.g.  ``is the value of this feature above or below 0.5?", see fig.~{\ref{twotrees}}). Each tree takes a randomly chosen subset of feature values as input, and returns a label prediction as output. For the data set we study in this paper there are in total 192 features, thus each decision tree will take as input values from a random subset of the 192 features. Following partitioning of the input data via splitting decisions, each individual tree will return a predicted value for k2.

Each tree in the RF ensemble is built directly from a `bootstrap sample' -- data  drawn at random from the training set, with each drawn example replaced prior to the subsequent selection. Therefore, individual examples can be represented more than once in the boot-strapped set. The final set contains the same number of examples as the original data set: typically one ends up sampling roughly two thirds of the original examples (due to the repeated selection of some examples). Examples not included in the boot-strapped data set used to fit a particular decision tree are called  ``out-of-bag" samples for that particular tree.

Following training, the trees can be used to predict labels for unseen examples of the same data type. Individual decision trees alone tend to over-fit training data and do not generalise well. RFs  solve this problem by fitting many trees on random sub-samples of the data (the ``boot-strapped"data) and then aggregating the results. By introducing randomness into the construction of each tree, a diverse set of models is produced, with the prediction of the entire ensemble given as the average of the predictions of the individual trees.

RFs can be used for both classification and regression tasks; here we will only be interested in regression. In particular, for our task we use `Extremely Randomized Trees', where the choice of threshold for splitting is randomised, and the best of these randomly chosen thresholds is used, rather than choosing the optimal threshold for splitting directly (as is the case for the standard approach, Random Forest Regressors). Extremely Randomised Trees are an example of ``weak" learners: although each individual tree may not fit the data very well, the averaged ensemble provides a good fit to the data that is likely to generalise.

The entire ensemble of decision trees is optimised according to a given metric. The most common optimality criterion for RF regression is the mean squared error (MSE), which we use here. Mean squared error loss penalises large discrepancies between predicted labels and true labels more than small ones:
\begin{equation}
MSE(y, \hat y) = \frac{1}{m}  \sum^{m-1}_{i=0} (y_i - \hat y_i)^2,
\end{equation}
where $y_i$ is the true label of the $i$-th example, $\hat y_i$ is the label predicted by the model and $m$ is the number of examples in the training data set.

Including the possibility of boot-strapping enables a measure of an `out-of-bag' (OOB) error. If each new tree is fit from a boot-strap sample of the training examples, $z_i = (x_i,y_i)$, then the OOB error is the average error for each $z_i$ calculated using predictions from those trees that \emph{do not} use $z_i$ in their boot-strap sample. One can either manually compare OOB error rates for particular choices of hyper-parameters, or perform an automated `grid-search' (which usually enables one to cover a larger range of hyper-parameters). Out-of-bag estimates are generally considered to provide accurate estimates of the generalisation error of the ensemble, often producing statistics that are even more precise, and computationally efficient than K-fold cross-validation estimates~\cite{Wolpert1999}.

The final, trained ensemble model itself is evaluated by the $R^2$ score, a metric which gives an indication of how well future examples are likely to be predicted by the model.  For a predicted value of $\hat y_i$ on the $i$-th example, with a true value of $y_i$, the $R^2$ score over $m$ samples is:
\begin{equation}\label{r2}
R^2(y,\hat y) = 1 - \frac{\sum^{m-1}_{i=0} (y_i-\hat y_i)^2}{\sum^{m-1}_{i=0} (y_i-\bar y)^2},
\end{equation}
where $\bar y$ is the mean: $ \frac{1}{m} \sum^{m-1}_{i=0} y_i$.

The best possible $R^2$ score is (arbitrarily close to) 1, and for extremely bad models can also be negative: models can be arbitrarily worse than simply estimating the mean. A model that is constant and always predicts the expected value of the labels, regardless of the training data, will have an $R^2$ of $0$ (the second term on the right-hand side of Eq~\ref{r2} simplifies to 1). In general one will measure the $R^2$ score for the training set, the validation set and the out-of-bag samples, in order to assess the overall quality of the model. A model with a high $R^2$ score for the training set, but a poor validation or OOB $R^2$ will be likely to generalise poorly. Ultimately, one aims for an OOB $R^2$ score as close to 1 as possible. The main advantage  of analysing the performance of the model on the validation set is to provide an opportunity to improve the model via hyper-parameter tuning, prior to final testing. Tuning the model runs the risk of over-fitting to the validation set, hence the need for a final test, or "hold out" set of data with which to measure the model's performance.

One faces a number of choices when implementing this kind of model. The number of trees, the maximum depth of each tree (how many splitting decisions are included), the minimum leaf number and the maximum features included in the root node will all contribute to the performance of the model. 

A final advantage of RFs is the possibility of assessing \emph{feature importance}. Individual features are ranked according to their contribution to the final model, providing an opportunity to prune redundant features and improve the computational efficiency of the final model.  Feature importance is evaluated \emph{after} the Random Forest has been trained: one randomly permutes all the example entries corresponding to a particular feature, and runs these altered examples through the model to generate a new set of predictions. The new predictions are then compared to the true labels to generate a new set of $R^2$ scores which can then be compared to those calculated prior to the random permutation. Any consequent reduction in $R^2$ score can be attributed to the particular feature being evaluated and all features can be ranked according to this measure.

\section{Methods}
\subsection{Data generation}\label{datagen}

Our aim was to simulate a system of fixed dimension  ($d=2$) and total number of time steps ($n=4$). Furthermore, we restrict our attention only to time-translation invariant processes. Therefore all the unitaries describing the joint system-environment evolution at different times are equal: $V_1=\dots = V_{n-1}=V$. This is a relatively natural assumption in many situations, where one does not expect a systematic change in the environment within each experimental run.

In several scenarios of practical interest, one would not expect the joint system-environment evolution to be completely arbitrary. In particular, for several quantum-technology applications one would attempt to have the system evolving as little as possible, aside from the chosen, controlled transformations. We therefore introduce a parameter $\phi\in [0,1]$ that describes how far from identity the unitary $V$ is. The latter is sampled in the following way: first a set of numbers $s_r$, $1\leq r \leq d$ is sampled independently and uniformly from the interval $[0,\phi]$. The diagonal matrix $D(s)= \textrm{diag}(e^{i 2 \pi s_1},\dots,e^{i 2 \pi s_d})$ is generated. Then, a $d$-dimensional unitary $G$ is sampled from the Haar measure, and finally $V=G D(s)G^{\dag}$ is calculated (we use Mezzadri's algorithm to sample unitary matrices from the Haar measure~\cite{mezzadri2006generate}). For $\phi=0$, this method only generates the identity matrix, while for $\phi=1$ it corresponds to sampling $V$ from the Haar measure. For intermediate values, $V$ applies random phases of at most $e^{i 2 \pi \phi}$ to the states of a randomly-chosen basis. We call $\phi$ the \emph{evolution} parameter. The initial joint state of system and non-Markovian environment is a $2k_2$-dimensional density matrix $\rho_0$ sampled from the `Ginibre' ensemble~\cite{BRUZDA2009320}.

A single example $x^i$ corresponds to a list of probabilities $p^i_{i_1,\dots,i_n}$, which is generated through formulas \eqref{step}, \eqref{dilated}, and \eqref{observed}. A set of values for the indices $(i_1,\dots,i_n)$, denotes a single feature. The value of the last index $i_n$,  denotes the basis of the final measurement, while the values of $i_j$, $1\leq j \leq n-1$, denote the unitary performed at time $t_j$. In our scenario, with $d=2$, the indices can vary in the range $0\leq i_j \leq 3$ for $1\leq j \leq n-1$ and $1\leq i_n \leq 3$. For $n=4$, this amounts to $4^{n-1}\times 3 = 4^{3}\times 3 = 192$ features. For example, the feature $(i_1,\dots,i_n)=(0,\dots,0,1)$ corresponds to the identity being applied at each time except the last, when the state $\ket{x_+}$ is measured. The full set of features also include the values of $k_1$, the `Markovian dimension', and the evolution parameter $\phi$, giving a total of $194$ features.
We generate examples for a range of parameter values ($k_1=1,2,4$; $k_2=1,2,8,16$; and $\phi=0.1,0.2,0.7,1$). For each parameter combination a total of $4096$ examples is generated, giving a total of $m= 196608$ examples.

The data set was shuffled and $70\%$ retained for training the RFs, with $20\%$ retained for hyper-parameter tuning. The final $10\%$ was retained for testing the tuned model. The $\log_2$ value of $k_2$ was chosen as the target feature for the network to learn.

A further test set, consisting of data generated using parameters outside those used to generate the training data was used to further assess the generalisability of the model.  Test set 1: $k_2=4$ with other parameters in the range $k_1=1,2,4$; $\phi=0.1,0.2,0.5,0.7,1$, with $256$ examples for each parameter combination. Test set 2: $k_1=2$; $k_2=64$; $\phi=0.5$. 

The data was generated using Mathematica software. The relevant notebooks can be obtained with permission from the authors.
\subsection{Learning algorithm}
The learning algorithm was implemented using code from the open-source python library Scikit-learn~\cite{pedregosa2011scikit}, specifically Scikit-learn's ExtraTreesRegressor (\href{http://scikit-learn.org/stable/modules/generated/sklearn.ensemble.ExtraTreesRegressor.html}{http://scikit-learn.org/stable/modules/generated/sklearn.ensemble.ExtraTreesRegressor.html}), and also from the open-source pytorch library: fastai (\href{https://github.com/fastai}{https://github.com/fastai}). The full code for the data simulation, pre-processing and model optimisation can be obtained with permission from the authors. Hyper-parameter tuning was initially optimised manually by comparing OOB $R^2$ estimates for a range of possible trial values. Optimal hyper-parameters values were assessed to be 80 trees, with all other values evaluated to agree with the Scikit-learn default values. Further hyper-parameter tuning was undertaken via both RandomCV search and GridSearch methods, although as the performance of the model did not change appreciably the Scikit-learn default values were retained.

The model was trained both including and excluding the parameters $\phi$ and $k_1$. Prior to training and testing the model, it was not clear to what extent these parameters were relevant to the learning task. This is the advantage of using simulated data to test a learning task: one can ask such questions even though the parameters may not be experimentally accessible. As we will see below, for this particular task, knowledge of the values of these parameters turns out to have little impact on the final accuracy of the trained model.

\section{Results}
In order to provide context, a dummy regression was performed on the training and validation data to calculate a comparative training $R^2$ score. In this way, one has a value of $R^2$ that is determined using a model that is simply calculating a ``naive" property of the data, such as the mean of the training labels. This gives a kind of benchmark with which to compare the $R^2$ scores obtained after training. A model designed to always predict the mean of the training set returned an $R^2$ score of $-9.9 \times 10^{-6}$, and a model designed to always predict a constant value of 2, independently of the input value, returned an OOB $R^2$ of $-9.6 \times 10^{-6}$. In both cases $R^2$ values are approximately 0. This is precisely what one would expect: recall that a model that can predict the expected value of the true labels, but does so independently of the training data, will have an $R^2$ of 0. Thus we expect $R^2$ scores following training and tuning of the Random Forests to be at least greater than 0, and ideally closer to 1.

Following hyper-parameter optimisation, training and validation of our model, feature importance was assessed in order to consider the possibility of training and testing with fewer input features. Model performance was assessed to remain stable until the total feature number reached 41 (out of a possible 194 features). Unsurprisingly, different features were assessed as optimal on different training runs due to the similarity of certain feature subsets. This is expected because each individual sequence of unitaries followed by a measurement should be as informative as any other sequence. 

Following feature reduction, the model was trained and validated on the reduced data set containing entries from these 41 features only, one of which was the $k_1$ value (somewhat surprisingly, the feature corresponding to $\phi$ was not ranked high enough to be included in the set of 41 features). Finally, the model was also trained and assessed after removing the $k_1$ feature. The results are contained in Table~\ref{fig1}.

\begin{table}[ht]%
\includegraphics[width=\columnwidth]{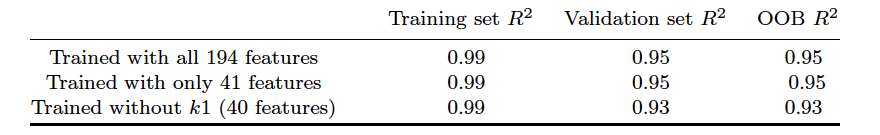}%
\caption{$R^2$ scores for $k_2$ prediction: validation set}
\label{fig1}%
\end{table}

Each of the three models was then evaluated using the hold-out set of test data, and the performance was seen to be essentially unchanged (Table~\ref{fig2}). The mean squared error of the $\log_2 k_2$ predicted by the model as compared to the true labels was also evaluated. By taking the square root of the MSE, one can assess the average deviation of the predicted values for $\log_2 k_2$ from the true values. These errors remained small for each of the three models (Table~\ref{fig2}, final column).

\begin{table}[ht]%
\includegraphics[width=\columnwidth]{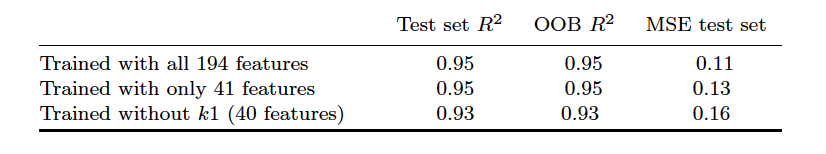}
\caption{$R^2$ scores for $k_2$ prediction: test set}
\label{fig2}
\end{table}

The results here indicate that the model would be expected to generalise reasonably well. The $R^2$ scores are a significant improvement over the values obtained from the naive dummy regressors. It is likely the small degree of over-fitting (evidenced by the discrepancy between training and validation $R^2$ scores, and the less-than-perfect OOB $R^2$) would be improved by training on larger data sets and with more trees. For many applications the total number of trees numbers in the thousands, where here we only used 80. The drawback of using such large numbers of trees is that training time will increase. For 80 trees training time was on the order of 30 seconds; for 200 trees it was on the order of two minutes with a corresponding improvement in OOB $R^2$ of 0.005. The results indicate a slight improvement (2\%) in performance when trained with knowledge of the $k_1$ parameter and no significant difference with adding knowledge of $\phi$.

Whilst the results suggest the model would generalise well to new data where parameters were consistent with those used to generate the data, further confidence was sought by testing on data generated according to parameters outside those used in simulation. We tested the model predictions for a true $\log_2 k_2$ of 2 across the full range of $k_1$ and $\phi$ values, including a previously unseen $\phi$ of 0.5. The results can be seen in Table~\ref{fig3} and show that although the model was not trained on examples generated for this particular value of $k_2$, the model was nonetheless able to provide a reasonably accurate prediction of unseen examples generated using this value of $k_2$. The mean prediction (bottom row of table) was close to $2$ in all three models

\begin{table}[ht]%
\includegraphics[width=0.75\columnwidth]{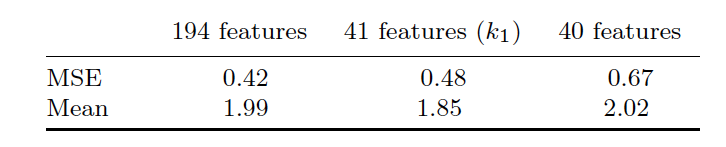}%
\caption{Model evaluation for test sets $log_2 k_2 = 2$}
\label{fig3}
\end{table}

We also tested the model on samples generated using a log$k_2$ of 6, and the mean of the predictions was $4.0$, with standard deviation of $0.1$. The result of 4.0 is as good as can be expected, given Random Forests can only return values within the training range. The MSE of $4$ simply reflects that the model is consistently predicting a $\log_2 k_2$ value of 4, which is being evaluated against the true  $\log_2 k_2$ value of $6$. These results are summarised in Table~\ref{fig4}.

\begin{table}[ht]%
\includegraphics[width=\columnwidth]{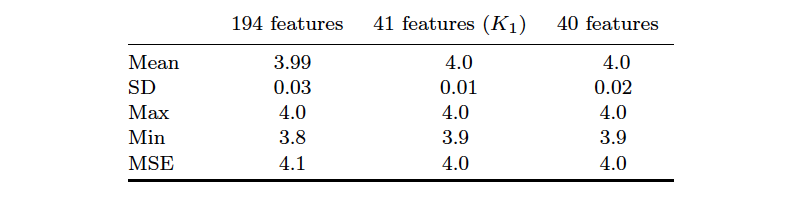}%
\caption{Model evaluation for test sets $log_2 k_2 = 6$}
\label{fig4}%
\end{table}

\section{Discussion}
Our results provide preliminary evidence that supervised learning techniques may well be useful for identifying features of non-Markovian noise in quantum experiments. Perhaps the most interesting result is that good predictions can still be made even when one considers tomographically incomplete data.

There are some obvious limitations of our approach. We are restricted to considering cases of fixed dimension and number of time-steps, and although it would be straightforward to extend the simulations to include variations of these parameters, clearly implementing the simulations will become increasingly onerous. An important question, in this respect, regards the scalability of the learning approach. A direct generalisation of our settings choices ($4$ unitaries per time step) would result in a total number of features (and thus of experimental settings) that grows exponentially with the number of  time steps (similarly to tomographic methods~\cite{giarmatzi2018quantum}). However, we have seen that there is a good amount of redundancy in the data, since retaining only $20\%$ of the features is sufficient to attain good accuracy. The question then is how the number of relevant features scales with the total number of time steps, and whether a less-than-exponential number would be sufficient. An independent question is whether a different choice of settings would be more efficient: here we used Pauli gates as a simple and natural choice but other choices might turn out to be more suitable, possibly depending on the class of processes under consideration.

There are many possibilities for future research, including testing alternative supervised learning methodologies. Deep neural networks, for example, may perform better on this particular regression task than the Random Forest technique we use here. It is impossible to know which kind of model architecture would perform best on this kind of data without directly comparing the performance of various architectures. Larger simulation sets, with more varied parameters, would also increase the likelihood that subsequently trained models would generalise beyond the limited setting we consider here. Most importantly, however, future research should be directed towards assessing to what degree these kinds of models can generalise to real world data. That is, data that has been generated experimentally rather than via simulation. Without further work, it is not possible to claim that our results would apply to real quantum experiments. Experimental platforms where one can accurately model Markovian and non-Markovian environments of specific dimensions would help validate the particular approach we take here.

\section*{Acknowledgments}
We are grateful to Josh Combes and Peter Wittek for helpful discussions. This work was supported by an Australian Research Council Centre of Excellence for Quantum Engineered Systems grant (CE 110001013). F.C.\ acknowledges support through an Australian Research Council Discovery Early Career Researcher Award (DE170100712).
This publication  was made possible through the support of a grant from the John Templeton Foundation. The opinions expressed in this publication are those of the authors and do not necessarily reflect the views of the John Templeton Foundation.

%%%%%%%%%%%%%%%%%%%%%%%%%%%
%%%%% BIBLIOGRAPHY %%%%%%%%
%%%%%%%%%%%%%%%%%%%%%%%%%%%

\bibliographystyle{linksen}
\bibliography{RandomForest}

\providecommand{\href}[2]{#2}\begingroup\raggedright\begin{thebibliography}{10}

\bibitem{biamonte2017}
J.~Biamonte, P.~Wittek, N.~Pancotti, P.~Rebentrost, N.~Wiebe, and S.~Lloyd,
  ``Quantum machine learning,'' {\em Nature} {\bfseries 549}, 195 (2017).

\bibitem{dunjko2017}
V.~Dunjko and H.~J. Briegel, ``Machine learning \& artificial intelligence in
  the quantum domain,'' \href{http://arxiv.org/abs/1709.02779}{{\ttfamily
  arXiv:1709.02779 [quant-ph]}}.

\bibitem{Costa2016}
F.~Costa and S.~Shrapnel, ``Quantum causal modelling,''
  \href{http://dx.doi.org/10.1088/1367-2630/18/6/063032}{{\em New Journal of
  Physics} {\bfseries 18}, 063032 (2016)}.

\bibitem{giarmatzi2018quantum}
C.~Giarmatzi and F.~Costa, ``A quantum causal discovery algorithm,''
  \href{http://dx.doi.org/10.1038/s41534-018-0062-6}{{\em npj Quantum
  Information} {\bfseries 4}, 17 (2018)}.

\bibitem{Pollock2018}
F.~A. Pollock, C.~Rodr\'{\i}guez-Rosario, T.~Frauenheim, M.~Paternostro, and
  K.~Modi, ``Operational Markov Condition for Quantum Processes,''
  \href{http://dx.doi.org/10.1103/PhysRevLett.120.040405}{{\em Phys. Rev.
  Lett.} {\bfseries 120}, 040405 (2018)}.

\bibitem{Pollock2018a}
F.~A. Pollock, C.~Rodr\'{\i}guez-Rosario, T.~Frauenheim, M.~Paternostro, and
  K.~Modi, ``Non-Markovian quantum processes: Complete framework and efficient
  characterization,'' \href{http://dx.doi.org/10.1103/PhysRevA.97.012127}{{\em
  Phys. Rev. A} {\bfseries 97}, 012127 (2018)}.

\bibitem{milz2016reconstructing}
S.~Milz, F.~A. Pollock, and K.~Modi, ``Reconstructing non-Markovian quantum
  dynamics with limited control,''
  \href{http://dx.doi.org/10.1103/PhysRevA.98.012108}{{\em Phys. Rev. A}
  {\bfseries 98}, 012108 (2018)}.
  \url{https://link.aps.org/doi/10.1103/PhysRevA.98.012108}.

\bibitem{Gutoski2006}
G.~Gutoski and J.~Watrous, ``Toward a general theory of quantum games,'' in
  {\em In Proceedings of 39th ACM STOC}, pp.~565--574.
\newblock 2006.
\newblock \href{http://arxiv.org/abs/quant-ph/0611234}{{\ttfamily
  quant-ph/0611234}}.

\bibitem{Chiribella2009}
G.~{Chiribella}, G.~M. {D'Ariano}, and P.~{Perinotti}, ``{Theoretical framework
  for quantum networks},''
  \href{http://dx.doi.org/10.1103/PhysRevA.80.022339}{{\em Phys. Rev. A}
  {\bfseries 80}, 022339 (2009)}.

\bibitem{Gross}
J.~A. Gross, C.~M. Caves, G.~J. Milburn, and J.~Combes, ``Qubit models of weak
  continuous measurements: markovian conditional and open-system dynamics,''
  \href{http://dx.doi.org/10.1088/2058-9565/aaa39f}{{\em Quantum Science and
  Technology} {\bfseries 3}, 024005 (2018)}.

\bibitem{Kretschmann2005}
D.~Kretschmann and R.~F. Werner, ``Quantum channels with memory,''
  \href{http://dx.doi.org/10.1103/PhysRevA.72.062323}{{\em Phys.\ Rev.\ A}
  {\bfseries 72}, 062323 (2005)}.

\bibitem{Chiribella2008}
G.~Chiribella, G.~M. D'Ariano, and P.~Perinotti, ``Quantum Circuit
  Architecture,'' \href{http://dx.doi.org/10.1103/PhysRevLett.101.060401}{{\em
  Phys. Rev. Lett.} {\bfseries 101}, 060401 (2008)}.

\bibitem{ACCARDI1978226}
L.~Accardi, ``Noncommutative Markov chains associated to a pressigned
  evolution: An application to the quantum theory of measurement,''
  \href{http://dx.doi.org/https://doi.org/10.1016/0001-8708(78)90012-9}{{\em
  Advances in Mathematics} {\bfseries 29}, 226 -- 243 (1978)}.

\bibitem{lindblad1979}
G.~Lindblad, ``Non-Markovian quantum stochastic processes and their entropy,''
  \href{http://dx.doi.org/10.1007/BF01197883}{{\em Comm. Math. Phys.}
  {\bfseries 65}, 281--294 (1979)}.

\bibitem{Allen2016}
J.-M.~A. Allen, J.~Barrett, D.~C. Horsman, C.~M. Lee, and R.~W. Spekkens,
  ``Quantum Common Causes and Quantum Causal Models,''
  \href{http://dx.doi.org/10.1103/PhysRevX.7.031021}{{\em Phys. Rev. X}
  {\bfseries 7}, 031021 (2017)}.

\bibitem{li2017concepts}
L.~Li, M.~J. Hall, and H.~M. Wiseman, ``Concepts of quantum non-Markovianity: a
  hierarchy,'' \href{http://arxiv.org/abs/1712.08879}{{\ttfamily
  arXiv:1712.08879 [quant-ph]}}.

\bibitem{Rivas2014}
{\'A}.~Rivas, S.~F. Huelga, and M.~B. Plenio, ``Quantum non-Markovianity:
  characterization, quantification and detection,''
  \href{http://dx.doi.org/10.1088/0034-4885/77/9/094001}{{\em Reports on
  Progress in Physics} {\bfseries 77}, 094001 (2014)}.

\bibitem{mitchell97}
T.~M. Mitchell {\em et al.}, ``Machine learning. WCB,'' 1997.

\bibitem{Breiman2001}
L.~Breiman, ``Random Forests,''
  \href{http://dx.doi.org/10.1023/A:1010933404324}{{\em Machine Learning}
  {\bfseries 45}, 5--32 (2001)}.

\bibitem{biau2016random}
G.~Biau and E.~Scornet, ``A random forest guided tour,'' {\em Test} {\bfseries
  25}, 197--227 (2016).

\bibitem{louppe2014}
G.~Louppe, ``Understanding random forests: From theory to practice,''
  \href{http://arxiv.org/abs/1407.7502}{{\ttfamily arXiv:1407.7502 [stat.ML]}}.

\bibitem{Caruana08}
R.~Caruana, N.~Karampatziakis, and A.~Yessenalina, ``An empirical evaluation of
  supervised learning in high dimensions,'' in {\em In International Conference
  on Machine Learning (ICML}, pp.~96--103.
\newblock 2008.

\bibitem{Wolpert1999}
D.~H. Wolpert and W.~G. Macready, ``An Efficient Method To Estimate Bagging's
  Generalization Error,'' \href{http://dx.doi.org/10.1023/A:1007519102914}{{\em
  Machine Learning} {\bfseries 35}, 41--55 (1999)}.

\bibitem{mezzadri2006generate}
F.~Mezzadri, ``How to generate random matrices from the classical compact
  groups,'' \href{http://arxiv.org/abs/math-ph/0609050}{{\ttfamily
  arXiv:math-ph/0609050}}.

\bibitem{BRUZDA2009320}
W.~Bruzda, V.~Cappellini, H.-J. Sommers, and K.~{\.{Z}}yczkowski, ``Random
  quantum operations,''
  \href{http://dx.doi.org/https://doi.org/10.1016/j.physleta.2008.11.043}{{\em
  Physics Letters A} {\bfseries 373}, 320 -- 324 (2009)}.

\bibitem{pedregosa2011scikit}
F.~Pedregosa, G.~Varoquaux, A.~Gramfort, V.~Michel, B.~Thirion, O.~Grisel,
  M.~Blondel, P.~Prettenhofer, R.~Weiss, V.~Dubourg, {\em et al.},
  ``Scikit-learn: Machine learning in Python,'' {\em Journal of machine
  learning research} {\bfseries 12}, 2825--2830 (2011).

\end{thebibliography}\endgroup

\end{document}